\documentclass[twocolumn,prb,aps,notitlepage]{revtex4-1}
\usepackage{tikz}
\usepackage{amssymb}
\usepackage{psfrag}
\usepackage{textcomp}
\usepackage{pdfsync}
\usepackage{amsmath}
\usepackage{amsfonts}
\usepackage{graphicx,bm,epstopdf}
\usepackage{subfigure}
\usepackage{comment}
\usepackage{verbatim}
\usepackage{color}
\usepackage{upgreek}
\usepackage{tikz}
\usepackage{natbib,hyperref}
\usepackage{threeparttable}
\usepackage{ifpdf}
\usepackage{ulem}

\begin{document}

\title{Theory of unidirectional spin Hall magnetoresistance in
heavy-metal/ferromagnetic-metal bilayers}
\author{Steven S.-L. Zhang}
\email{zhangshule@missouri.edu}
\author{Giovanni Vignale}
\affiliation{Department of Physics and Astronomy, University of Missouri, Columbia, MO
65211}
\date{\today }

\begin{abstract}
Recent experiments have revealed nonlinear features of the magnetoresistance
in metallic bilayers consisting of a heavy-metal (HM) and a ferromagnetic
metal (FM). A small change in the longitudinal resistance of the bilayer has
been observed when reversing the direction of either the applied in-plane
current or the magnetization. We attribute such nonlinear transport behavior
to the spin-polarization dependence of the electron \textit{mobility} in the
FM layer acting in concert with the spin accumulation induced in that layer
by the spin Hall current originating in the bulk of the HM layer. An
explicit expression for the nonlinear magnetoresistance is derived based on
a simple drift-diffusion model, which shows that the nonlinear
magnetoresistance appears at the first order of spin Hall angle (SHA), and
changes sign when the current is reversed, in agreement with the
experimental observations. We also discuss possible ways to control the sign of the nonlinear magnetoresistance and to enhance the magnitude of effect.
\end{abstract}

\maketitle

\affiliation{Department of Physics and Astronomy, University of Missouri, Columbia, MO
65211}

Recently, nonlinear magnetoresistance has been observed experimentally in
several heavy metal (HM)/ferromagnetic metal (FM) bilayers~\cite%
{Avci15NatPhys_unidirectionAMR,ajFerguson16APL,kjLee16arxiv}. In these
experiments, both the current and the magnetization of the FM layer lie in
the plane of the layers and are mutually perpendicular, as shown in Fig.~\ref%
{fig:schematics}. For a given magnitude of the current density (in the range
of $10^{7}\sim 10^{8}$ $A/$cm$^{2}$), it has been found that the longitudinal
resistance changes when the current direction is reversed. Furthermore, by
injecting an a.c.~current, it has been observed that the 2nd harmonic
component of the longitudinal resistance changes sign as the magnetization
direction is reversed~\cite{Avci15NatPhys_unidirectionAMR}: this shows
that, different from the familiar linear transport, the magnetoresistance
has a linear dependence on the current density.

A definitive interpretation of these experimental observations has not yet
emerged. Avci et al. associated the nonlinear magnetoresistance with the modulation of interface scattering potential induced by the spin Hall effect and the ensuing interfacial resistance change, similar to the interfacial contribution of giant magnetoresistance (GMR)~\cite{Valet93PRB,Bass1999JMMM_review-CPP-GMR,Parkin93PRL_gmr-interface,Falicov92PRB}. Another interpretation of the effect invokes magnon excitation in
the FM layer due to electron spin-flip scattering at the interface~\cite%
{ajFerguson16APL,kjLee16arxiv}. While this process has recently been shown
to play a key role in the spin-charge conversion in
HM/ferromagnetic-insulator (FI) layered structures~\cite%
{sZhang12PRL,sZhang12PRB,Kajiwara10Nature,jShi16natComm_magnon-drag,vanWees15NatPhys,xfHan16PRB}%
, it is usually neglected in metallic systems, for which it is a good
approximation to assume that the spin current is continuous at the interface~%
\cite{Valet93PRB}.

\begin{figure}
\includegraphics[width=\linewidth]{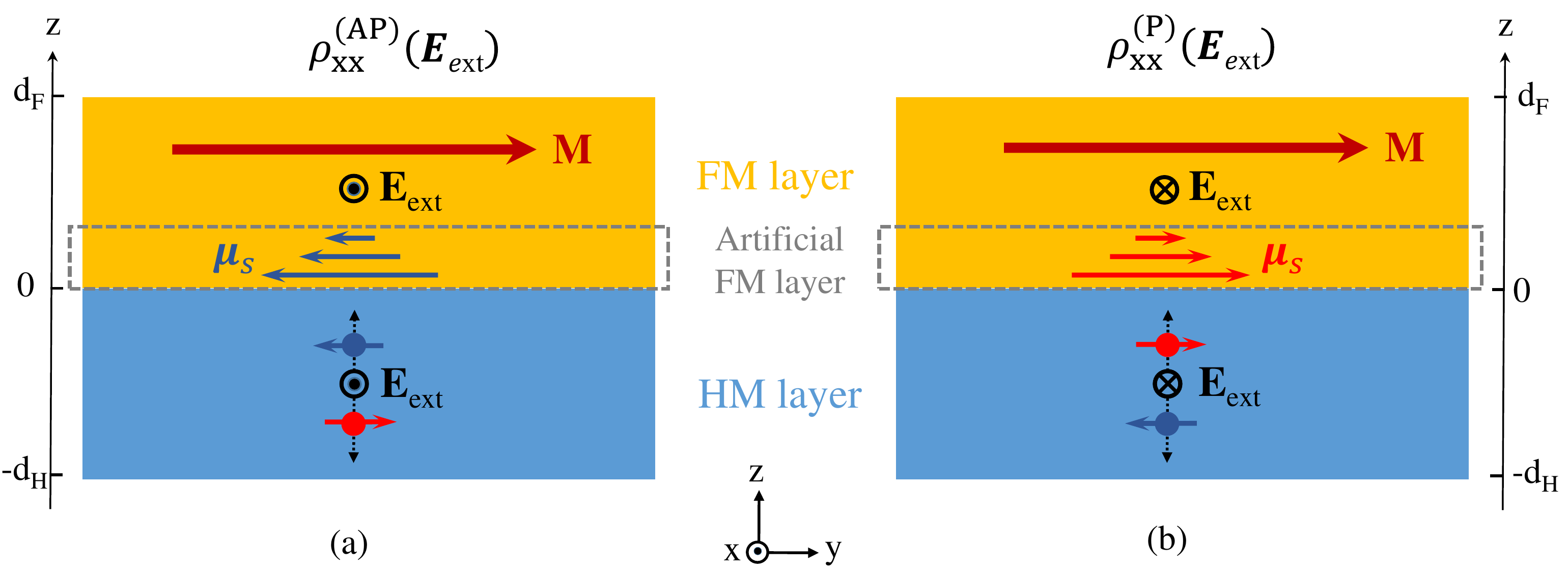}
\caption{Schematic diagrams showing the mechanism of nonlinear longitudinal
resistivity in HM/FM bilayers due to spin accumulation in the FM layer
induced by spin Hall effect in the HM layer. The external electric field $%
\mathbf{E}_{ext}$ is applied in the positive $x$-direction in (a) and in the
negative $x$-direction in (b). The dotted arrows in the HM layer denotes the
directions of the spin Hall currents. The solid arrows in the grey dashed
boxes describe the magnitude and direction of the spin accumulation $%
\boldsymbol{\protect\mu}_s$ which may be regarded as an artificial FM layer.
The difference in longitudinal resistivity of the bilayer in antiparallel (a) and parallel (b) configurations
arises from the spin-dependence of
the mobility in the FM layer in analogy with CIP-GMR. }
\label{fig:schematics}
\end{figure}

In this paper, we present a simple analytical theory of the nonlinear
magnetoresistance in HM/FM bilayers. We propose that the effect arises from
the combined action of the spin accumulation induced by the spin Hall effect
in the HM layer and the spin-polarization dependence of the electron
mobility in the FM layer. As schematically shown in Fig.~\ref{fig:schematics}%
, when an in-plane current is driven in a HM/FM bilayer, a spin Hall current
flowing perpendicular to the layers is generated in the bulk of the HM layer
and subsequently creates spin accumulation on both sides of the interface.
Spin accumulation is known as a local quantity that characterizes an excess
density of electrons with one specific spin orientation and a corresponding
depletion of electrons with the opposite spin orientation, so that no charge
accumulation is created. Although such local spin dependent density
variation in HM layer would not alter the conductivity of the layer in which
the mobility of electrons is spin-independent, the conductivity of the FM
layer is indeed modified by the spin accumulation inside the layer. This may
be best understood by thinking of the spin accumulation near the interface
as an artificial ferromagnetic layer. Based on our understanding of the
current-in-plane (CIP) GMR~\cite{Fert88PRL,Grunberg89PRB,camley89PRL,Levy90PRL}, we would anticipate a change in longitudinal
resistance when the ``magnetization" of the artificial FM layer (i.e., the
direction of the spin accumulation) switches from parallel to antiparallel
(or vice versa) to that of the ``natural" FM layer. The only difference from
CIP-GMR lies in the fact that ``magnetization" of the artificial FM layer is
generated by the electric current itself via the spin transport
perpendicular to the layers. This simple analogy immediately demonstrates
the nonlinear character of the corresponding magnetoresistance effect.

When an in-plane current is applied to a HM/FM bilayer, a spin current
propagating perpendicular to the layers is generated by the spin Hall effect~%
\cite{Hirsch99PRL,sZhang00PRL,Vignaleg10JSNM,Sinova15RMP_SHE} in the HM.
This transverse spin current affects the linear in-plane resistivity via the
inverse spin Hall effect -- a phenomenon that has been intensively studied~\cite{Chien12PRL_Proximity-Pt,Althammer13PRB_SH-MR,Wees13PRB_SH-MR,Hahn13PRB_SMR-Pt-Ta,slZhang16PRL,Casanova16PRB-SMR,Junyeon16PRL_SMR-metallic}
and goes under the name of ``spin-Hall magnetoresistance"~\cite%
{Saitoh13PRL_SH-MR,Bauer13PRB_SH-MR}. In addition, the modulation of the
electron spin density in the ferromagnet generates a nonlinear resistivity
as discussed in the introduction and shown in detail below. Here we treat
both linear and nonlinear contributions on equal footing through a set of
equations that couple the spin and charge transport in directions parallel
and perpendicular to the plane of the layers.

To be specific, let us assume the external electric field is applied in the $%
x$-direction, i.e., $\mathbf{E}_{ext}=E_{x}\hat{\mathbf{x}}$ ($E_{x} $ could
be either positive or negative), and fix the magnetization vector of the FM
layer in the positive $\hat{\mathbf{y}}$ direction which is also taken as
the quantization axis of the electron spin. The HM/FM interface is located
at $z=0$. The general drift-diffusion equation for electrons with spin
orientation $\alpha $ ($\alpha =\pm $ or $\uparrow \left( \downarrow \right)
$ denoting the spin orientation parallel or antiparallel to the
magnetization) can be written as follows

\begin{equation}
j_{x}^{\alpha }\left( z\right) =\sigma ^{\alpha }\left( z\right)
E_{x}-\alpha \theta j_{z}^{\alpha }\left( z\right)
\label{Eq: j_x^alpha-general}
\end{equation}%
and
\begin{equation}
j_{z}^{\alpha }\left( z\right) =\sigma ^{\alpha }\left( z\right) \frac{d}{dz}%
\mu ^{\alpha }\left( z\right) +\alpha \theta j_{x}^{\alpha }\left( z\right)\,,
\label{Eq: j_z^alpha-general}
\end{equation}%
where $j_{i}^{\alpha}$ is the current density carried by spin-$\alpha$
electrons with $i=x$ or $z$ denoting the spatial direction of flow, $\theta $
is the bulk SHA and $\mu ^{\alpha }\left( z\right) $ is the spin-dependent
chemical potential, which is related to the nonequilibrium part of the
electron density $n^{\alpha }\left( z\right) $ as follows
\begin{equation}
\mu ^{\alpha }\left( z\right) =\left[ N^{\alpha }\left( \varepsilon
_{F}\right) \right] ^{-1}n^{\alpha }\left( z\right) -\phi \left( z\right)\,,
\label{Eq: mu^a(z)-general}
\end{equation}%
with $N^{\alpha }\left( \varepsilon _{F}\right) $ being the density of
states of spin-$\alpha$ electrons at Fermi level, and $\phi \left( z\right) $
being the spin-independent part of the chemical potential. Notice that in
Eqs.~(\ref{Eq: j_x^alpha-general}) and (\ref{Eq: j_z^alpha-general}) we have
assumed a spatially dependent local conductivity controlled by the electron
spin density as follows:
\begin{equation}
\sigma ^{\alpha }\left( z\right) =\nu ^{\alpha }\left[ n_{0}^{\alpha
}+n^{\alpha }\left( z\right) \right]\,,  \label{Eq: sigma^a(z)-general}
\end{equation}%
where $n_{0}^{\alpha }$ and $\nu ^{\alpha }$ are the equilibrium density and
the mobility of spin-$\alpha$ electrons, respectively. The charge and $y$-spin current densities are defined as 
$j_{i}\left( z\right) \equiv j_{i}^{\uparrow }\left( z\right)
+j_{i}^{\downarrow }\left( z\right)$
and
$Q_{i}^{y}\left( z\right) \equiv j_{i}^{\uparrow }\left( z\right)
-j_{i}^{\downarrow }\left( z\right)$.

We also assume that
charge neutrality is locally maintained, i.e.,
\begin{equation}
n^{\uparrow }\left( z\right) +n^{\downarrow }\left( z\right) =0\,.
\label{Eq: charge neutrality}
\end{equation}%
In metals, this is justified by the observation that the integrated space
charge vanishes beyond a very short screening length -- of the order of
Angstroms. This supplemental condition links the transport in the two spin
channels. In what follows, we shall discuss the transport in each layer separately.

In the HM layer, the equilibrium conduction electron density is
spin-independent and therefore transport coefficients such as the mobility
and the diffusion constant can be taken to be spin-independent up to
first-order in the current-induced spin polarization. Equations~(\ref{Eq:
j_x^alpha-general}) and (\ref{Eq: j_z^alpha-general}) reduce to

\begin{equation}
j_{x}^{\alpha }\left( z\right) =\sigma _{H}^{\alpha }\left( z\right)
E_{x}-\alpha \theta _{H}j_{z}^{\alpha }\left( z\right)
\label{Eq: j_x^alpha-H-full}
\end{equation}%
and
\begin{equation}
j_{z}^{\alpha }\left( z\right) =\frac{1}{2}\alpha \theta _{H}\sigma
_{0,H}E_{x}+\sigma _{0,H}\frac{d}{dz}\mu ^{\alpha }\left( z\right)\,,
\label{Eq: j_z^alpha-H-1stOrder}
\end{equation}%
where $\sigma _{0,H}=\nu _{H}n_{0,H}$ is the bulk Drude conductivity of the
HM. Notice that in the second of these equations we are keeping only terms up to
first order in $\theta _{H}$.

In steady state, the spin dependent current density satisfies the
generalized continuity equation
\begin{equation}
\frac{d}{dz}j_{z}^{\alpha }\left( z\right) =\sigma _{0,H}\frac{d^{2}}{dz^{2}}%
\mu ^{\alpha }\left( z\right) =\frac{n^{\alpha }\left( z\right) -n^{-\alpha
}\left( z\right) }{\tau _{sf,H}}\,,  \label{Eq: continty-eqn-j_z^a-H}
\end{equation}%
where $\tau _{sf,H}$ is the spin-flip relaxation time. With Eq.~(\ref{Eq:
mu^a(z)-general}), we may express the right hand side of this equation in
terms of the chemical potentials which are found to satisfy the following
differential equations
\begin{equation}
\frac{d^{2}}{dz^{2}}\mu _{c}\left( z\right) =0  \label{Eq: mu+&mu-_homo}
\end{equation}%
and
\begin{equation}
\frac{d^{2}}{dz^{2}}\mu _{s}\left( z\right) -\frac{\mu _{s}\left( z\right) }{%
L_{H}^{2}}=0\,,  \label{Eq: mu+&mu-_inhomo}
\end{equation}%
where we have defined the sum and difference of the chemical potentials as $%
\mu _{c}\left( z\right) \equiv \left[ \mu ^{\uparrow }\left( z\right) +\mu
^{\downarrow }\left( z\right) \right] /2$ and $\mu _{s}\left( z\right)
\equiv \left[ \mu ^{\uparrow }\left( z\right) -\mu ^{\downarrow }\left(
z\right) \right] /2$ respectively, and $L_{H}\equiv \sqrt{\sigma _{0,H}\tau
_{sf,H}/2N_{H}\left( \varepsilon _{F}\right) }$ as the spin diffusion length.

For the transport in the FM layer, we neglect the anomalous Hall effect
since the SHA is usually an order of magnitude smaller than that in the HM
layer. This assumption allows us to simplify the equations for current
densities in FM layer as
\begin{equation}
j_{x}^{\alpha }\left( z\right) =\nu _{F}^{\alpha }\left[ n_{0,F}^{\alpha
}+n^{\alpha }\left( z\right) \right] E_{x}  \label{Eq:j_x^alpha(z)_FM}
\end{equation}%
and
\begin{equation}
j_{z}^{\alpha }\left( z\right) =\sigma _{0,F}^{\alpha }\frac{d}{dz}\mu
^{\alpha }\left( z\right)\,,  \label{Eq: j_z^alpha(z)_FM}
\end{equation}%
with $\sigma _{0,F}^{\alpha }$ being the bulk conductivity of the spin$%
-\alpha $ channel in the FM layer. In steady state, the continuity equation
reads%
\begin{equation}
\frac{d}{dz}j_{z}^{\alpha }\left( z\right) =\sigma _{0,F}^{\alpha }\frac{%
d^{2}}{dz^{2}}\mu ^{\alpha }\left( z\right) =\frac{n^{\alpha }\left(
z\right) -n^{-\alpha }\left( z\right) }{\tau _{sf,F}}\,. \label{Eq: dj/dz-FM}
\end{equation}%
Making use of Eqs.~(\ref{Eq: mu^a(z)-general}) and (\ref{Eq: charge
neutrality}) to express $n^{\alpha } -n^{-\alpha}$ in terms of $\mu^{\alpha
} -\mu^{-\alpha}$
we find that the equation for $\mu_c$ takes the form
\begin{equation}
\frac{d^{2}}{dz^{2}}\mu _{c}\left( z\right) +p_{\sigma }\frac{d^{2}}{dz^{2}}%
\mu _{s}\left( z\right) =0\,,  \label{Eq: diff-eq_mu+-_FM}
\end{equation}%
where $p_{\sigma}\equiv\left( \sigma _{0,F}^{\uparrow }-\sigma
_{0,F}^{\downarrow }\right) /\left( \sigma _{0,F}^{\uparrow }+\sigma
_{0,F}^{\downarrow }\right) $ is the conductivity spin asymmetry. On the
other hand, the equation for $\mu_s$ remains of the same form as in the HM
layer (Eq.~(\ref{Eq: mu+&mu-_inhomo}) except for replacing $L_H$ by the
ferromagnetic spin diffusion length
$L_{F}=\sqrt{\sigma _{0,F}\left( 1-p_{\sigma }^{2}\right) \tau
_{sf,F}/2N_{F}\left( \varepsilon _{F}\right) \left( 1-p_{N}^{2}\right) }$,
where $p_{N}\equiv\left( N_{F}^{\uparrow }-N_{F}^{\downarrow }\right)
/\left( N_{F}^{\uparrow }+N_{F}^{\downarrow }\right) $ is the spin asymmetry
in the density of states at Fermi level.

For the boundary conditions at the interface ($z=0$), neglecting interfacial
spin-flip scattering and the small interfacial resistance \cite%
{Valet93PRB,jBass07JPhysCondMatt_spin-flip}, we assume that both the spin
current density flowing in $z$-direction and the chemical potentials are
continuous, i.e.,
$Q_{z}^{y}\left( 0^{-}\right) =Q_{z}^{y}\left( 0^{+}\right) $ and $\mu
^{\alpha }\left( 0^{-}\right) =\mu ^{\alpha }\left( 0^{+}\right) $. At the
same time, since there is no charge flow in the $z$-direction, we set $%
j_{z}\left( z\right) =0$ everywhere. Also, we take $Q_{z}^{y}$ to vanish at
the two outer surfaces, i.e., at $z=-d_{H}$ for the HM layer and $z=d_{F}$
for the FM layer with $d_{H}$ and $d_{F}$ being the thicknesses of the HM
and FM layers respectively.

By inserting the general solutions of the chemical potentials and the spin
current densities into the boundary conditions for each interface, we can
now determine all transport quantities of interests.

For example, up to first order in $\theta_H$, the in-plane charge current
density in the FM layer is given by 
\begin{equation}
j_{x}\left( z\right) =\sigma _{0,F}E_{x}+\frac{1}{2}\left( \nu
_{F}^{\uparrow }-\nu _{F}^{\downarrow }\right) \left[ n^{\uparrow }\left(
z\right) -n^{\downarrow }\left( z\right) \right] E_{x}\,,  \label{Eq: j_x(z>0)}
\end{equation}%
where $\sigma _{0,F}=n_{0,F}^{\uparrow }\nu _{F}^{\uparrow
}+n_{0,F}^{\downarrow }\nu _{F}^{\downarrow }$ is the total bulk
conductivity of the FM and the spin accumulation is given by
\begin{widetext}
\begin{equation}
n^{\uparrow }\left( z\right) -n^{\downarrow }\left( z\right) =-\frac{2\left(
\theta _{H}L_{H}\right) N_{F}\left( \varepsilon _{F}\right) \left(
1-p_{N}^{2}\right) \tanh \left( \frac{d_{H}}{2L_{H}}\right) \cosh \left(
\frac{d_{F}-z}{L_{F}}\right) }{\cosh \left( \frac{d_{F}}{L_{F}}\right)
+\left( 1-p_{\sigma }^{2}\right) \left( \frac{\sigma _{0,F}L_{H}}{\sigma _{0,H}L_{F}}\right)
\sinh \left( \frac{d_{F}}{L_{F}}\right) \coth \left( \frac{d_{H}}{L_{H}}\right)}E_{x}\,.
\label{Eq: n^up-n^dn_sol}
\end{equation}
\end{widetext}
Note that the negative sign in front of the expression on the r.h.s. of
Eq.~(\ref{Eq: n^up-n^dn_sol}) implies that minority electrons are
accumulated near the interface when both $\theta _{H}$ and $E_{x}$ are
positive.

Equations~(\ref{Eq: j_x(z>0)}) and (\ref{Eq: n^up-n^dn_sol}) are quite
remarkable. Firstly, we observe that the correction to the in-plane charge
current density (i.e., the second term on the r.h.s.~of Eq.~(\ref{Eq:
j_x(z>0)})) is proportional to $E_{x}^{2}$, since $n^{\uparrow }\left(
z\right) -n^{\downarrow }\left( z\right) $ is the linear response of the
spin density to the external electric field. Secondly, the nonlinear
contribution appears at the first order in the SHA, in contrast to the
linear spin Hall magnetoresistance which is known to be of second order in
the SHA~\cite{Bauer13PRB_SH-MR}. The above features qualitatively agree with
recent experimental observations~\cite%
{Avci15NatPhys_unidirectionAMR,Avci15APL_AMR}.

Equation~(\ref{Eq: j_x(z>0)}) makes clear that, in our interpretation, the
spin dependence of the electron \textit{mobility}, i.e., the nonzero value
of $\left( \nu_{F}^{\uparrow }-\nu _{F}^{\downarrow }\right)$, is essential
to the appearance of a nonlinear magnetoresistance. Indeed, if the
mobilities were not spin-dependent, the total in-plane conductivity $\sigma
^{\uparrow}+\sigma ^{\downarrow }$ would remain unchanged by virtue of the
charge neutrality condition~(\ref{Eq: charge neutrality}). This is exactly
what happens in the HM layer, where the in-plane charge current remains
unchanged up to $O\left( \theta _{H}\right) $. The underlying physics is
rather transparent: If majority electrons in the FM layer exhibit higher
mobility than minority electrons (i.e., $\nu _{F}^{\uparrow }>\nu
_{F}^{\downarrow }$), then accumulation of majority electrons will lead to
an increase in the conductivity, and vice versa. The crucial role of spin asymmetry in the electron mobility of the FM is also consistent with the absence of nonlinear magnetoresistance effect in HM/FI bilayers (such as Pt/YIG) measured in recent experiments~\cite{Avci15APL_AMR}.

The total longitudinal resistivity of the bilayer can be calculated as
$\rho _{xx}=(d_H+d_F)E_{x}/{\int dzj_{x}\left( z\right)}$
where the current density is integrated over the thickness of the bilayer. Similar to GMR,
the amplitude of the unidirectional spin Hall magnetoresistance (USMR) is
characterized by the ratio

\begin{equation}
USMR=\frac{\rho _{xx}\left( E_{x}\right) -\rho _{xx}\left( -E_{x}\right) }{%
\rho _{xx}\left( E_{x}\right) }\,.  \label{Eq: UMR-def}
\end{equation}
Up to first order in $\theta_H$, we obtain

\begin{widetext}
\begin{equation}
USMR\simeq 6\left( \frac{\sigma _{0,F}L_{F}}{\sigma _{0,H}d_{H}+\sigma
_{0,F}d_{F}}\right) \frac{\left( p_{\sigma }-p_{N}\right) \left( \theta
_{H}E_{x}L_{H}/\varepsilon _{F}\right) \tanh \left( \frac{d_{H}}{2L_{H}}%
\right) \tanh \left( \frac{d_{F}}{L_{F}}\right) }{1+\left( 1-p_{\sigma
}^{2}\right) \left( \frac{\sigma _{0,F}L_{H}}{\sigma _{0,H}L_{F}}\right) \tanh \left( \frac{%
d_{F}}{L_{F}}\right) \coth \left( \frac{d_{H}}{L_{H}}\right)}\,,
\label{Eq: NLMR-1stOrder}
\end{equation}
\end{widetext}
where we have used the relations $\nu _{F}^{\alpha }=\sigma _{0,F}^{\alpha
}/n_{0,F}^{\alpha }$ and $N_{F}^{\alpha }=3n_{0,F}^{\alpha }/2\varepsilon
_{F}$ for free electron model with $\varepsilon _{F}$ being the Fermi energy
of the FM. Note that $\nu _{F}^{\uparrow }-\nu _{F}^{\downarrow }$ is
proportional to the difference of the spin polarization of conductivity $%
p_{\sigma }$ and of density of states at Fermi energy $p_{N}$.

\begin{figure}
\includegraphics[width=0.8\linewidth]{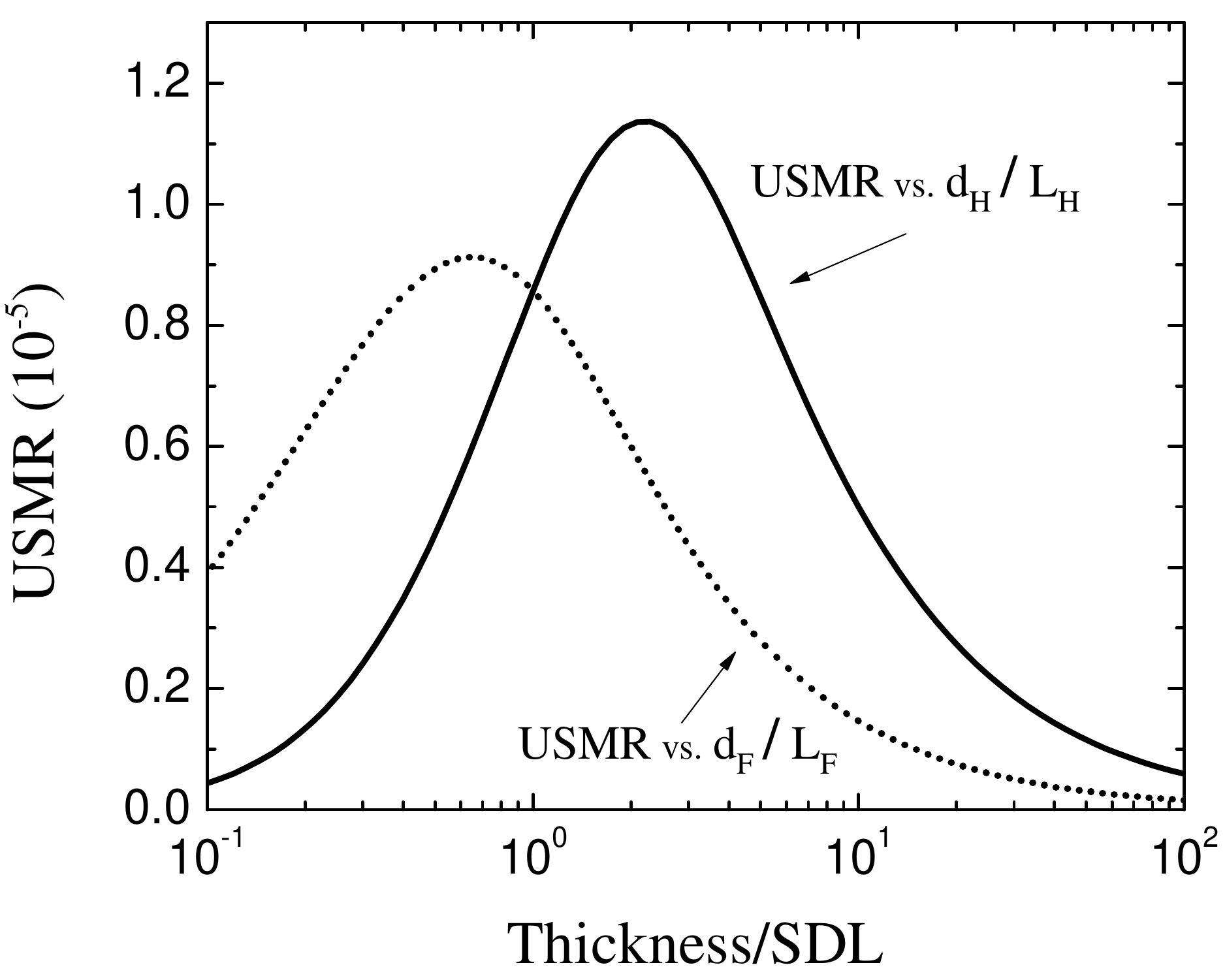}
\caption{USMR as a function of the thickness of the HM layer (scaled with
the spin diffusion length (SDL) $L_{H} $) for fixed thickness of the FM
layer of $d_{F}=L_{F}=10$ nm (solid line) and as a function of the thickness
of the FM layer (scaled with $L_{F}$) for fixed thickness of the HM layer of
$d_{H}=L_{H}=5$ nm (dotted line). Other parameters assumed in the numerical
calculation are $\protect\theta_H=0.1$, $\left| {E_x}\right|=10^{-4}$ $V$%
/nm, $p_{\sigma}-p_N=0.5$, $\protect\epsilon_F=5$ $eV$ and $\protect%
\sigma_{0,H}=\sigma_{0,F}=0.033$ $(\mu \Omega~\mathrm{cm}%
)^{-1}$. The calculations are done to first order in SHA.}
\label{fig:thick-dep}
\end{figure}

In Fig.~\ref{fig:thick-dep}, we plot the USMR as a function of the thickness
of one layer while the thickness of the other is fixed. As we have pointed
out previously, although the leading order nonlinear correction to the
in-plane current density only occurs in the FM layer, the HM layer also
plays an essential role by inducing the spin accumulation in the FM layer
via the spin Hall effect. Therefore, when the thickness of either layer
becomes much smaller than the corresponding spin diffusion length, the USMR
diminishes. On the other hand, when the
thickness of either layer is much larger than the spin diffusion length,
more current is shunted into the bulk of the layers and hence the
interfacial effect of USMR gets diluted as indicated by the prefactor on the
r.h.s. of Eq.~(\ref{Eq: NLMR-1stOrder}). Not surprisingly, the USMR peaks
around the respective spin diffusion length of each layer. The dependence of
the USMR on the thickness of the HM layer agrees qualitatively with
experiments, whereas the dependence on the thickness of the FM layer has not yet been measured.

With Eq.~(\ref{Eq: NLMR-1stOrder}), we can also make quantitative
comparisons of the calculated magnitude of USMR with experimentally observed
values. For a Pt (6 nm)/Co (3 nm) bilayer with $E_{x}=10^{-4}~$V/nm
and the following material parameters: $\theta _{H}=0.1$~\cite%
{lqLiu12PRL_SH-switchinng,Parkin15NatPhys,Avci15NatPhys_unidirectionAMR}, $%
\sigma _{H}=0.02$ $\left( \mu \Omega~\mathrm{cm}\right) ^{-1}$, $L_{H}=5$
nm, $\sigma _{F}=0.05$ $\left( \mu \Omega~\mathrm{cm}\right) ^{-1}$, $%
L_{F}=40$ nm~\cite{jBass07JPhysCondMatt_spin-flip}, $\varepsilon _{F}=5$ $eV$,
$p_{\sigma }-p_{N}=0.5$, we find USMR $\simeq 0.9\times 10^{-5}$, which is
only a factor of $2$ smaller than the experimental value~\cite%
{Avci15NatPhys_unidirectionAMR,Avci15APL_AMR}. We have checked
that the USMR is negligibly reduced in the presence of an interfacial resistance $r_{I}\sim 1$%
\textbf{\ }f$\Omega$~m~\cite{Bass14JMMM_Co-Pt}.

We notice that the magnitude of the USMR may be underestimated due to
several simplifying assumptions we adopted in our model calculation. First,
in the derivation of Eq.~(\ref{Eq: NLMR-1stOrder}), we assumed spherical
Fermi surfaces and constant density of states at Fermi energy. Strong energy
dependence of the density of states near the Fermi surface (e.g., in Ni~\cite%
{Aziz09PRL}) may enhance the effect just as it enhances the spin
accumulation-induced nonlinear GMR effect observed in dual spin valves~\cite%
{Aziz09PRL}. Second, we neglected the indirect influence of the spin
accumulation on transport parameters such as the bulk and interfacial
resistances due to electron-electron correlation or shift of scattering
potential.

In the presence of interfacial spin-flip scattering, there would be a partial loss of spin current across the interface (known as spin memory loss~\cite{jBass07JPhysCondMatt_spin-flip,Sanchez14PRL-spin-memory-loss,kChen15PRL_spin-pumping,Kovalev16ArXiv_spin-memory-loss,Bass2016JMMM_cpp-GMR-review}).  This effect can be easily incorporated in our treatment through a simple change in the boundary conditions for the spin current.~\cite{UMR-footnote}  Spin memory loss, treated in this way,  results in a reduction of the USMR given in Eq.~(\ref{Eq: NLMR-1stOrder}) by a factor of order unity.  However, the absorbed spin current may in turn lead to additional contribution to the USMR via interfacial spin-dependent scattering~\cite{Tokura16PRL_UMR-TI}.

An interesting observation based on Eq.~(\ref{Eq: NLMR-1stOrder}) is that
the USMR depends linearly on the difference of $p_{\sigma }-p_{N}$, which
suggests that the sign of the USMR also depends on the overall sign of $%
p_{\sigma }-p_{N}$. In Ref.~\cite{Fert76JPhysF} Fert and Campbell showed
that the signs of \thinspace $p_{\sigma}$ for various binary alloys of
transition metals may change depending on the relative position of the
d-bands of the host and the impurity. For example, they showed that $p_{\sigma }$ of NiFe is positive whereas that
of FeCr is negative~\cite{Fert76JPhysF}. By making use of this property, a
``reversed" CIP-GMR, that is to say, a CIP-GMR in which the antiparallel
alignment of magnetizations has \textit{lower} resistance than the parallel
arrangement, could be explained in a Fe/Cu super-lattice with half of the Fe
layers being intercalated with thin Cr layers~\cite{Fert94PRL}. Similar
experiments can be carried out to test our theory of USMR. For example, we
predict that the USMR in Pt/FeCr should have opposite sign than that in
Pt/NiCr bilayer.

Our model calculation also suggests several ways to enhance the USMR. For
metallic systems, the effect would be amplified in an asymmetric trilayer
structure of the form HM$_{1}$/FM/HM$_{2}$ with HM$_{1}$ and HM$_{2}$ having
opposite signs of $\theta_H$ (for example, HM$_{1}$=Pt and HM$_{2}$=Ta). In
such a structure, the orientations of the spin accumulations on opposite
sides of the FM layer will be identical, hence the contributions of the two
interfaces to the USMR will add constructively. Our theory also suggests
that an enhanced USMR may be found in paramagnetic and ferromagnetic
semiconductor bilayers, which have much lower carrier densities than their
metallic counterparts. As shown by Eq.~(\ref{Eq: NLMR-1stOrder}), the USMR
is inversely proportional to the Fermi energy which scales with the
equilibrium free electron density as $\varepsilon _{F} \propto n_e^{2/3}$.
Very recently, Olejn\'{\i}k \textit{et al.} found that the USMR in
ferromagnetic-(Ga,As)Mn/paramagnetic-(Ga,As)Mn bilayers is larger than that
in metallic bilayers by several orders of magnitudes~\cite%
{Jungwirth15PRB_nonlinear-MR-semicond}, and they attributed the big
enhancement to the low carrier densities in their semiconducting systems.

In summary, we have developed a drift-diffusion theory for HM/FM bilayers
with an in-plane electric current. The theory is self-consistent in the
sense that it takes into account the effect of the current-induced spin
accumulation on the longitudinal resistance. The unidirectional magnetoresistance is an effect of first order in the spin
Hall angle of the HM layer, in contrast to the linear spin Hall
magnetoresistance which is an effect of second order in the spin Hall angle.
We have suggested ways to control the sign of the nonlinear
magnetoresistance and to amplify the magnitude of the effect by judicious
choice of materials and/or nanostructure engineering. It appears that
conducting bilayers consisting of a ferromagnet and a paramagnetic metal
with large spin Hall angle have considerable potential to work as reversible
diodes that may be controlled by the magnetic direction of the ferromagnetic
layer.

It is a pleasure to thank A.~Fert, P.~Gambardella, O.~Heinonen, E.~Saitoh, S.~Zhang,
and W.~Zhang for stimulating discussions. This work was supported by NSF
Grants DMR-1406568.

\bibliographystyle{my-asp-style}
\bibliography{20160521_bilayer-GMR}

\begin{thebibliography}{45}
\bibitem [{UMR()}]{UMR-footnote}%
  \BibitemOpen
  \href@noop {} {}\bibinfo {note} {Spin memory loss can be conveniently
  introduced into our formulation through interfacial boundary condition of the
  spin current as $Q_z^y(0^+)=\eta Q_z^y(0^-)$, where the phenomenological
  parameter $\eta$ varies from 0 to 1, characterizing the interfacial spin
  memory loss ($\eta$ = 0 when the spin current injected from the HM layer is
  completely absorbed by the interface, and $\eta$ = 1 when no absorption of
  spin current takes place at the interface). With this boundary condition, one
  can check that the resulting USMR is approximately equal to the one given by
  Eq. (18) multiplied by $\eta$.}\BibitemShut {Stop}%
\end{thebibliography}

\end{document}